\documentclass[12pt]{article}
 
\usepackage{setspace}

\usepackage[utf8]{inputenc} 
\usepackage[T1]{fontenc}    
\usepackage{hyperref}       
\usepackage{url}            
\usepackage{booktabs}       
\usepackage{amsfonts}       
\usepackage{nicefrac}       
\usepackage{microtype}      
\usepackage{graphicx} 
\usepackage{float}  
\usepackage{subfigure} 
\usepackage{mathtools}
\usepackage{amsmath}
\usepackage{esvect}
\usepackage{caption}

\title{The Influence of Nuisance Parameter Uncertainty on Statistical Inference in Practical Data Science Models}

\author{
  Yunrong Wan\\
  Department of Applied Mathematics and Statistics\\
  Johns Hopkins University\\
  \texttt{ywan13@jh.edu} \\
  JHU 553.809 Master's Research \\
  Final Report\\
}
\begin{document}
\maketitle
\begin{abstract}
For multiple reasons -- such as avoiding overtraining from one data set or because of having received numerical estimates for some parameters in a model from an alternative source -- it is sometimes useful to divide a model's parameters into one group of primary parameters and one group of nuisance parameters. However, uncertainty in the values of nuisance parameters is an inevitable factor that impacts the model's reliability. This paper examines the issue of uncertainty calculation for primary parameters of interest in the presence of nuisance parameters. We illustrate a general procedure on two distinct model forms: 1) the GARCH time series model with univariate nuisance parameter and 2) multiple hidden layer feed-forward neural network models with multivariate nuisance parameters. Leveraging an existing theoretical framework for nuisance parameter uncertainty, we show how to modify the confidence regions for the primary parameters while considering the inherent uncertainty introduced by nuisance parameters. Furthermore, our study validates the practical effectiveness of adjusted confidence regions that properly account for uncertainty in nuisance parameters. Such an adjustment helps data scientists produce results that more honestly reflect the overall uncertainty.


\end{abstract}
\begin{spacing}{1.5}
\section{Introduction}
The process of mathematical model building typically involves parameter estimation. To optimize computational efficiency in terms of both calculation quantity and time, researchers sometimes categorize parameters that demand substantial time for calculation within the model but are not the focus of the study as nuisance parameters. These parameters are then assigned values based on empirical knowledge to streamline the calculation process. Despite being considered of secondary importance, especially in data science modeling with a large number of parameters, the values and uncertainties in the nuisance parameters have an impact on subsequent downstream analysis, possibly leading to invalid estimates of primary parameters and even wrong conclusions (Carroll \cite{carroll2003variances}). Therefore, the impact of nuisance parameter uncertainty on main parameters has become a point that needs attention in the creation of data science models.

The study of nuisance parameters has been an ongoing research topic.
Basu~\cite{basu2011elimination} reviewed nuisance parameter elimination methods, evaluating marginalization and conditioning, offering a Bayesian perspective. Van der Laan \cite{vanderLaan+2014+29+57} emphasized estimating treatment-specific means while addressing nuisance parameters using data-adaptive estimators, introducing targeted minimum loss-based estimators (TMLEs) for effective bias reduction and propensity score targeting in statistical inference. Elliott et al. \cite{elliott2015nearly} explored nonstandard hypothesis testing problems involving a nuisance parameter, establishing an upper bound on the weighted average power of valid tests and presenting a numerical algorithm for identifying feasible tests. Andrews and Mikusheva \cite{AndrewsandMikusheva2016conditional} introduced a novel Gaussian conditional test to address nuisance parameters in moment equality models without identification assumptions, presenting a sufficient statistic and constructing conditional tests to effectively handle nuisance parameters in weakly identified models. Spall \cite{spall1989effect} presented a result for the asymptotic distribution and the primary parameters variance while taking account of the uncertainty in the nuisance parameters.
The estimated primary variance, incorporating information from various data sources regarding nuisance parameters, leads to more accurate confidence intervals for primary parameters. This approach, rooted in maximum likelihood estimation, is general enough to find practical applications in diverse fields. For example, Spall and Garner \cite{spall1990parameter} show how the method accurately reflects the impact of nuisance parameters in state-space models in aerospace.

Maximum likelihood estimation (MLE) is a statistical method employed to estimate the primary parameters, denoted as a vector $\boldsymbol{\theta}$, of a probabilistic model that best explain the observed data. Parameters of secondary importance, but still essential for obtaining accurate estimates of $\boldsymbol{\theta}$, are denoted as a vector $\boldsymbol{\alpha}$ and are commonly referred to as nuisance parameters. The MLE framework defines a likelihood function that quantifies the probability of observing the given data under different parameter values, and then identifies the parameter values that maximize this likelihood function, often symbolized as $L(\boldsymbol{\theta}$). It is often more convenient to work with the logarithm of the likelihood function, denoted as log($L(\boldsymbol{\theta}$)), as it simplifies computations and retains the same maxima. The resulting estimates are the MLEs that offer a point estimate for the true values of the parameters, exhibiting desirable asymptotic properties and playing a fundamental role in statistical inference. The precision of the parameter estimates is measured by the Fisher information matrix (FIM) whose inverse is used to calculate the covariance matrix associated with the MLE.

This article applies Spall's theory to three distinct data science models: 1) nonlinear (exponential) regreesion, 2) the GARCH time series model and 3) multiple hidden layer feedforward neural network, situated in the realms of finance and machine learning, respectively. The objective is to investigate how the uncertainty surrounding nuisance parameters affects relevant broader measures of uncertainty. The general approach of this paper may apply to other models as well. 

The remainder of the report is organized as follows: Section 2 presents the theoretical methodology; Section 3 assesses the method's accuracy in the context of a nonlinear regression model; Section 4 discusses the application models; Section 5 presents the numerical studies; and Section 6 summarizes the results.


\section{\large General Nuisance Parameters Theory Methodology}

This section presents the main result on asymptotic normality from \cite{spall1989effect}. The theory is grounded in the asymptotic distribution of parameters. The main result pertains to asymptotic normality of the scaled/centered estimate of $\boldsymbol{\theta}$ in the presence of uncertainty in the chosen value of $\boldsymbol{\alpha}$. Let $\boldsymbol{X}^{(n)}$ denote the sample of size $n$ that will be used to estimate $\boldsymbol{\theta}$, $\hat{\boldsymbol{\alpha}}^{(m)}$ denote the given $\boldsymbol{\alpha}$ used to estimate $\boldsymbol{\theta}$, and $m$ denote the size of information used to get the value of  $\hat{\boldsymbol{\alpha}}^{(m)}$ from an independent (different) resource.

Basing on maximum likelihood estimation of $\boldsymbol{\theta}$ given $\boldsymbol{X}^{(n)}$ and $\hat{\boldsymbol{\alpha}}^{(m)}$, the estimation of $\boldsymbol{\theta}$, denoted by $\hat{\boldsymbol{\theta}}$, is expressed as (\ref{eq:1}):

\begin{equation}
    \hat{\boldsymbol{\theta}}(\boldsymbol{X}^{(n)},\hat{\boldsymbol{\alpha}}^{(m)})=\{ \boldsymbol{\theta}: \boldsymbol{s}(\boldsymbol{\theta}|\boldsymbol{X}^{(n)},\hat{\boldsymbol{\alpha}}^{(m)})=\boldsymbol{0} \} , \label{eq:1}
\end{equation}
where $\boldsymbol{s}( \cdot )\equiv\partial {\rm log}L / \partial \boldsymbol{\theta}$, $L$ is the likelihood function for $\hat{\boldsymbol{\theta}}$ given $(\boldsymbol{X}^{(n)}$,$\hat{\boldsymbol{\alpha}}^{(m)})$.
Then we need $\hat{\boldsymbol{\theta}}'\equiv \partial \hat{\boldsymbol{\theta}} / \partial \boldsymbol{\alpha}^T$
, which can be expressed in term of $\boldsymbol{s}( \cdot )$, the score function (gradient) of the log-likelihood function, by the implicit function theorem as:

\begin{equation}
     \hat{\boldsymbol{\theta}}'(\boldsymbol{X}^{(n)}, \boldsymbol{\alpha})=-\left(\frac{\partial \boldsymbol{s}}{\partial \boldsymbol{\theta}^T}\right)^{-1}\frac{\partial \boldsymbol{s} }{\partial \boldsymbol{\alpha}^T},  \label{eq:2}
\end{equation}
where it is assumed that the indicated (matrix) gradients and matrix inverse exist. And $\hat{\boldsymbol{\theta}}''(\boldsymbol{X}^{(n)}, \boldsymbol{\alpha}) = \partial^2\hat{\boldsymbol{\theta}}/\partial\boldsymbol{\alpha}^T\partial\boldsymbol{\alpha}^T$ is given in \cite{spall1989effect}.  There are two conditions (Spall \cite{spall1989effect}) for the main theorem. We let $\boldsymbol{\theta}^*$ and $\boldsymbol{\alpha}^*$ denote the true (unknown) values of $\boldsymbol{\theta}$ and $\boldsymbol{\alpha}$ respectively.

\textbf{\textit{Condition 1}}: For all $n$ sufficiently large and almost all $\boldsymbol{X}^{(n)}$ (with respect to the probability measure $P(\boldsymbol{X}^{(n)}|\boldsymbol{\alpha}^*,\boldsymbol{\theta}^*)$), $\hat{\boldsymbol{\theta}}''(\boldsymbol{X}^{(n)}, \boldsymbol{\alpha})$ exists and is continuous in an open region about $\boldsymbol{\alpha}^*$.

\textbf{\textit{Condition 2}}: As $ n \rightarrow \infty $,
\begin{equation}
    \hat{\boldsymbol{\theta}}'(\boldsymbol{X}^{(n)}, \boldsymbol{\alpha}^*) \xrightarrow{\rm pr.} \boldsymbol{D}_1, \label{cond:2.1}
\end{equation}
\begin{equation}
    \hat{\boldsymbol{\theta}}''(\boldsymbol{X}^{(n)}, \boldsymbol{\alpha}^*) \xrightarrow{\rm pr.} \boldsymbol{D}_2,\label{cond:2.2}
\end{equation}
where pr. represents convergence in probability and $\boldsymbol{D}_1$ and $\boldsymbol{D}_2$ are constant matrices.

Let Conditions 1 and 2 hold. And since we mentioned above that $\boldsymbol{X}^{(n)}$ and $\hat{\boldsymbol{\alpha}}^{(m)}$ are from different (independent) data sources, $\boldsymbol{X}^{(n)}$ and $\hat{\boldsymbol{\alpha}}^{(m)}$ are statistically independent, and we know:
\begin{equation}
    \hat{\boldsymbol{\alpha}}^{(m)} \stackrel{\rm a.d}{\sim} N\left( \boldsymbol{\alpha}^*, \frac{\boldsymbol{V}_{\boldsymbol{\alpha}}}{m}\right), m\rightarrow\infty,
\end{equation}
\begin{equation}
    \hat{\boldsymbol{\theta}}(\boldsymbol{X}^{(n)}, \boldsymbol{\alpha}^*) \stackrel{\rm a.d}{\sim} N\left( \boldsymbol{\theta}^*, \frac{\boldsymbol{V}_{\boldsymbol{\theta}}}{n}\right), n\rightarrow\infty,
\end{equation}
where a.d means asymptotically distributed , $\boldsymbol{V}_{\boldsymbol{\alpha}}/m$ is the per-sample variance for $ \hat{\boldsymbol{\alpha}}^{(m)}$, and $\boldsymbol{V}_{\boldsymbol{\theta}}/n$ is the per-sample variance for $\hat{\boldsymbol{\theta}}$, which empirically is the inverse average FIM in MLE of $\boldsymbol{\theta}$ given $\boldsymbol{\alpha}^*$. Assuming that $\boldsymbol{X}^{(n)}$ and $\hat{\boldsymbol{\alpha}}^{(m)}$ are independent, the main result from the Spall \cite{spall1989effect} theorem can be expressed as follows:

\textbf{\textit{Theorem}}: As $n\rightarrow\infty$, $m\rightarrow\infty$,
\begin{equation}
    \hat{\boldsymbol{\theta}}(\boldsymbol{X}^{(n)},\hat{\boldsymbol{\alpha}}^{(m)}) \stackrel{\rm a.d}{\sim} N\left( \boldsymbol{\theta}^*, \frac{\boldsymbol{V}_{\boldsymbol{\theta}}}{n}+\boldsymbol{D}_1\frac{\boldsymbol{V}_{\boldsymbol{\alpha}}}{m}\boldsymbol{D}_1^T\right), \label{theorem}
\end{equation}
where $\boldsymbol{D}_1$ is the limiting value of $\hat{\boldsymbol{\theta}}'$. Expression (\ref{theorem}) represents the asymptotic distribution of the estimated primary parameters, accounting for the uncertainty of the nuisance parameters. Notably, when no nuisance parameters are present, the usual asymptotic normality of $\hat{\boldsymbol{\theta}}$ is preserved.

This theorem finds practical application in real-world modeling by providing insights into the variance of the estimated primary parameters. In practical scenarios, the quantity $\boldsymbol{V}_{\boldsymbol{\theta}}/n$ can be estimated by taking the inverse of the FIM. 

Estimation or approximation of $\boldsymbol{V}_{\boldsymbol{\alpha}}/m$ in (\ref{theorem}) is crucial in order to make the Theorem useful in applications. The optimal scenario involves performing MLE of the nuisance parameter using an independent sample and then calculating $\boldsymbol{V}_{\boldsymbol{\alpha}}/m$ from its own FIM, which is the methodology applied below to the GARCH model.

If a separate MLE process is not used for $\alpha$, an alternative approach involves a Bayesian-type method to determine $\hat{\boldsymbol{\alpha}}^{(m)}$ and to approximate $\boldsymbol{V}_{\boldsymbol{\alpha}}/m$. 
This type of process is used in the context of neural network applications (see Section 3.2 below) and in the aerospace application in Spall and Garner \cite{spall1990parameter}

\section{\large Numerical Test of Nuisance Parameter Adjustment}
To evaluate the accuracy of the variance adjustments in (\ref{theorem}), we compare it with the variance derived from estimating all parameters by following a nonlinear regression. The purpose of this study is to compare the nuisance parameter adjustment in (\ref{theorem}) with the uncertainty that would be obtained if all parameters ($\boldsymbol{\theta},\boldsymbol{\alpha}$) were estimated from one dataset (which is difficult or infeasible in some cases). 

Initially, an exponential model ($y \sim {\rm exp}(\lambda))$ is formulated as expression (\ref{eq:exp}):
\begin{equation}
    f(y_i | x_i, \boldsymbol{\theta},\alpha) = \lambda_i e^{-\lambda_i y_i}, \\
    \lambda = e^{b_0 + b_1 x_1 + b_2 x_2}
,\label{eq:exp}
\end{equation}
where primary parameters $\boldsymbol{\theta} = (b_0, b_2)$ and scalar nuisance parameter $\alpha = b_1 $, serving as the basis for generating two independent datasets through random simulation. Dataset A, comprising $n=$1000 observations, is dedicated to estimating primary parameters, while dataset B, consisting of $m=$50 observations, is utilized for estimating nuisance parameters alongside their variances. This deliberate choice aims to accentuate the estimation error associated with nuisance parameters given the limited sample size.

Subsequently, we explore four distinct scenarios using this dataset:  in scenario 1, primary parameters are estimated as MLE using dataset A following the acquisition of nuisance parameter estimates from dataset B. The unadjusted variance of the primary parameters is yielded as the inverse of the FIM. In scenario 2, building upon this, we integrate the nuisance parameter variances from dataset B into the primary parameters estimation process in dataset A, resulting in the adjusted variance of primary parameters shown in (\ref{theorem}). In scenario 3, it exclusively relies on dataset A to estimate all parameters.  Primary parameters variances are extracted from the variance matrix of all parameters by utilizing the inverse of the FIM of MLE. In scenario 4, we combine datasets A and B into a unified dataset, followed by the same procedure as scenario 3 to gain primary parameter variances. In addition to the unadjusted variance in scenario 1, the remaining three scenarios all incorporate the uncertainty of nuisance parameters into the estimation process.

For each estimated primary parameters pair's value and its corresponding variance, we construct a 90\% confidence ellipse to check whether the true value lies within the ellipse. Subsequently, the experiment is replicated with a fresh set of randomly generated datasets, repeating the process 1000 times. Finally, the ratio of the 90\% confidence ellipses containing the true values for the four variances is computed, providing a comprehensive assessment of the accuracy of our variance adjustments. The result are shown below:
\begin{figure}[H]
\centering
\includegraphics[ width=1\textwidth]{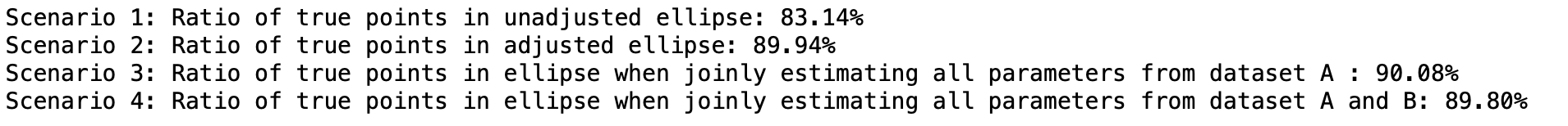} 
\captionsetup{font=small}
\label{Fig.expo1} 
\end{figure}

Also, to provide a visualization of the results, we selected the datasets pair whose Euclidean distance between the estimated values and true values in scenario one is the median value of all tests as our sample. We then plotted four 90\% confidence ellipses along with the true value to illustrate the situation graphically. The result is shown in Figure. \ref{Fig.expo2}:
\begin{figure}[H]
\centering
\includegraphics[ width=0.9\textwidth]{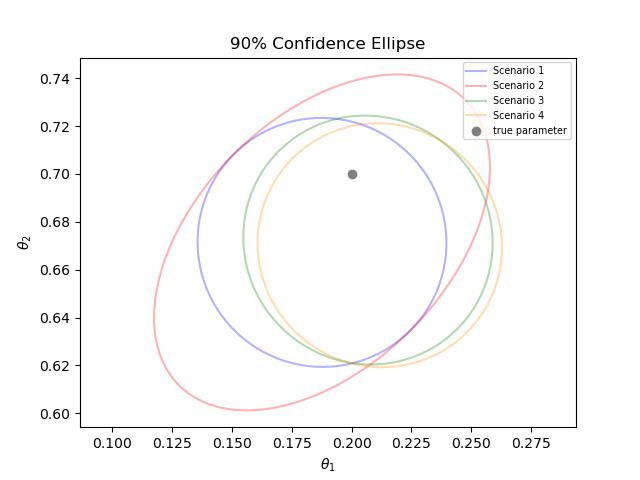} 
\captionsetup{font=small}
\caption{90\% confidence ellipses and true value $\boldsymbol{\theta}$ in median distance case.}
\label{Fig.expo2} 
\end{figure}

 Basing on the ratios and the plot, the ratios of the 90\% confidence ellipses covering the true values for variances in scenario 2--4 are all approximately near 90\%, while unadjusted one is smaller than 90\% (approx 83\%). Therefore, our adjusted variance, compared to the unadjusted one, enables more accurate and reliable statistical inferences. This is, the nuisance parameter method in (\ref{theorem}) provides the correct level of probability coverage, as we set out to verify.

\section{Models}
This section describes the time series (GARCH) and neural network models that we consider in our study of the effect of nuisance parameters. Section 5 will present the associated numerical results
\subsection{GARCH model}
In financial markets, the return rate stands out as one of the most crucial indicators. Among the various models employed to scrutinize return rates, the generalized autoregressive conditional heteroskedasticity (GARCH) model (Bollerslev et al. \cite{bollerslev1986generalized}) developed from the autoregressive conditional heteroskedasticity (ARCH) model (Engle \cite{engle1982autoregressive}), is widely recognized as a preeminent statistical model.

Let us introduce the standard representation of the GARCH model first. Let $y_t$ denote the return on the asset of interest at time $t, t=1,...,T$, $\mu_t$ denote the expected value of $y_t$ given information of time $t-1$, $\epsilon_t$ be a discrete time stochastic process, serially uncorrelated with mean zero and conditional variance $\sigma_t^2$, which may be changing through time. Thus, $\sigma_t^2$ is a time series composed of the ARCH terms ($a_i\sigma_{t-i}^2$), the GARCH terms ($b_i\epsilon_{t-i}^2$), and a constant term ($w$). In this model, $a_i$ and $b_i$ are primary parameters, while $w$ is a nuisance parameter. Equivalently, $y_t$ can be expressed as follows: 
\begin{equation}
    y_t=\mu_t+\epsilon_t ,  \label{eq:garch_y}
\end{equation}
\begin{equation}
    \epsilon_t=z_t\sigma_t,  \label{eq:garch_epsilon}
\end{equation}
\begin{equation}
    \sigma_t^2 = \omega + \sum_{i=1}^q a_i \epsilon_{t-i}^2 + \sum_{i=1}^p b_i \sigma_{t-i}^2 , \label{eq:garch}
\end{equation}
where $z_t$ are i.i.d, $\mathbf{E}(z_t)=0$, ${\rm var}(z_t)=1$, $q$ is the order of the ARCH terms and $p$ is the order of the GARCH terms. According these equation, it is  clear that $\mu_t=\mathbf{E}(y_t|y_{t-1},\epsilon_{t-1},z_{t-1})$, ${\rm var}(y_t|y_{t-1},\epsilon_{t-1},z_{t-1})={\rm var}(\epsilon_t|y_{t-1},\epsilon_{t-1},z_{t-1})=\sigma_t^2$.
The GARCH model is focusing on $\sigma_t^2$, the conditional variance of $y_t$, which is also the conditional variance of $\epsilon_t$. This was pioneered by Bollerslev \cite{bollerslev1986generalized} with a flexible lag structure, and extensively employed for analyzing and modeling time-varying volatility in financial markets. Serving as the cornerstone of financial time series models, the GARCH model has given rise to a diverse array of derivative models, including but not limited to EGARCH (exponential GARCH), IGARCH (integrated GARCH), and others. Consequently, the GARCH model holds significant importance and research value within the realm of financial analysis.

This paper mainly focuses on the GARCH(1,1) model specified as below, which stands with one lag in time series term:
\begin{equation}
    \sigma_t^2 = \omega + a \epsilon_{t-1}^2 + b \sigma_{t-1}^2, \label{eq:garch_small}
\end{equation}
where $\omega>0$, $a\geq0$,$b\geq0$ and $a+b<1$, which ensures $\epsilon_t$ is covariance stationary (Bollerslev et al. \cite{BOLLERSLEV19925}). A time series is considered covariance stationary if its mean, variance, and autocorrelation structure do not change over time. In the GARCH context, such stationarity is essential because it allows for the stability of statistical properties over time, making the model more reliable and suitable for forecasting.

In the GARCH model, the ARCH term($ a \epsilon_{t-1}^2$) captures the short-term impact of past volatility, and the GARCH term($b \sigma_{t-1}^2$) reflects the longer-term persistence in volatility. The terms allow the GARCH model to adapt and update volatility estimates based on recent information. Thus, $a$ and $b$ are the primary parameters in the GARCH(1,1) model that people are interested in. The constant term $\omega$ is often considered as a nuisance parameter which is assigned the value 0 (Liu and Brorsen \cite{liu1995maximum}) or an extremely small value to improve the efficiency of the primary parameters estimation. However, the impact of constant term's uncertainty on the model is still worthy of concern. Pakel et al. \cite{pakel2011nuisance} used composite likelihood to weaken the impact of the nuisance parameter's uncertainty on estimation. In contrast, we include the uncertainty of the nuisance parameters in the approximation of parameter variance by MLE-based methodology, thereby obtaining a more honest confidence interval.

\subsubsection{Model structure}

Applying the general nuisance parameters theory methodology into the GARCH model, the estimated primary parameters are found according to: (\ref{eq:garch_1}):
\begin{equation}
(\hat{a}(\boldsymbol{y}^{(n)},\hat{\omega}^{(m)}),\hat{b}(\boldsymbol{y}^{(n)},\hat{\omega}^{(m)}))^T=\{ a,b: \boldsymbol{s}(a, b|\boldsymbol{y}^{(n)},\hat{\omega}^{(m)})=\boldsymbol{0} \}.  \label{eq:garch_1}
\end{equation}
Expression (\ref{theorem}) is developed as (\ref{eq:garch_2}):
\begin{equation}
    (\hat{a}(\boldsymbol{y}^{(n)},\hat{\omega}^{(m)}),\hat{b}(\boldsymbol{y}^{(n)},\hat{\omega}^{(m)})^T \stackrel{a.d}{\sim} N\left( (a^*,b^*)^T,{\rm cov}((\hat{a},\hat{b})^T)\right), \label{eq:garch_2}
\end{equation}
where cov$((\hat{a},\hat{b})^T)$, $ \boldsymbol{V}_{a,b}$, and $\boldsymbol{D}_1$ is expressed in detail below:
\begin{equation}
    {\rm cov}((\hat{a},\hat{b})^T)={\rm cov}( \hat{\boldsymbol{\theta}})=\frac{\boldsymbol{V}_{a,b}}{n}+\boldsymbol{D}_1\frac{\boldsymbol{V}_{\omega}}{m}\boldsymbol{D}_1^T , \label{eq:garch_3}
\end{equation}
\begin{equation}
    \boldsymbol{V}_{a,b}= {\rm cov}((\hat{a},\hat{b})^T|\hat{\omega}^*), \label{eq:garch_4}
\end{equation}
\begin{equation}
    \boldsymbol{D}_1=-\left(\frac{\partial \boldsymbol{s}}{\partial \boldsymbol{\theta}^T}\right)^{-1}\frac{\partial \boldsymbol{s} }{\partial \omega}, \label{eq:garch_5}
\end{equation}
where $\boldsymbol{\theta}=(a,b)^T$. Then $-\left(\partial \boldsymbol{s} / \partial \boldsymbol{\theta}^T \right)^{-1}$ is the inverse of the negative Hessian matrix of the log-likelihood function, whose expectation is equal to the expectation of the inverse of FIM of MLE $\boldsymbol{\theta}$ given $\omega^*$, which is the conditional variance of $(\hat{a},\hat{b})^T$  ($\boldsymbol{V}_{a,b}$). $\boldsymbol{V}_{a,b}$ can be used to estimate $-\left(\partial \boldsymbol{s} / \partial \boldsymbol{\theta}^T \right)^{-1}$. Therefore, we can rewrite (\ref{eq:garch_3}) as follows:
\begin{equation}
    {\rm cov}((\hat{a},\hat{b})^T|\boldsymbol{y}^{(n)},\hat{\omega}^{(m)}) \approx \frac{\boldsymbol{V}_{a,b}}{n}+\boldsymbol{V}_{a,b}\left(\frac{\partial \boldsymbol{s} }{\partial \omega}\right)\frac{\boldsymbol{V}_{\omega}}{m}\left(\frac{\partial \boldsymbol{s} }{\partial \omega}\right)^T\boldsymbol{V}_{a,b}^T .\label{eq:garch_var}
\end{equation}

After specifying the general theory above, the specific calculations proceed below. The full conditional probability density of $y_t$ is assumed normal: 
\begin{equation}
    f(y_t|\sigma_t,\mu_t) = (2\pi\sigma_t^2)^{-\frac{1}{2}} \cdot {\rm exp}\left(-\frac{(y_t - \mu_t)^2}{2\sigma_t^2}\right) .\label{eq:garch_detail_1}
\end{equation}
In the estimation of the GARCH model, the technique of variance targeting has been introduced from Engle \cite{engle2002dynamic} in order to eliminate the conditional heteroscedasticity. The parameters in the model can be estimated while adhering to the variance targeting constraint, which means the expectation of the unconditional variance can be assumed as in expression (\ref{eq:variance-targeting}). This constraint ensures that the long-run variance-covariance matrix matches the sample covariance matrix:
\begin{equation}
    \mathbf{E}(\sigma_t^2)=\frac{\omega}{1-a-b} ,   \\ \mathbf{E}(y_t)=0. \label{eq:variance-targeting}
\end{equation}
As in the long term or large number of sampling, the GARCH model goes into the steady state which implies that the parameters have converged to values where the conditional volatility is no longer changing significantly from one period to the next. The volatility will not be affected by the previous information. When $y_t$ be a strictly stationary GARCH(1,1) model (Williams \cite{williams2011Garch}), then
\begin{equation}
    \sigma_t^2(\boldsymbol{\theta})=\frac{\omega}{1-b} +\sum_{k=1}^\infty ab^{k-1}y^2_{t-k} ,\label{eq:gl1}
\end{equation}
where $\boldsymbol{\theta}=(a,b)$ are primary parameters and $\sigma_0^2(\theta)=\omega/(1-a-b)$. For $n\rightarrow\infty, \boldsymbol{\theta}_n\xrightarrow{\rm a.s}\boldsymbol{\theta}_0 $, where a.s means "almost surely" and $\boldsymbol{\theta}_0$ is the true value of $\boldsymbol{\theta}$. The log-likelihood function (Williams \cite{williams2011Garch}) is:
\begin{equation}
    {\rm log}(L) = \frac{n-1}{2}{\rm log}(2\pi)+\frac{1}{2}\sum_{t=1}^n({\rm log}\sigma^2_t(\boldsymbol{\theta})+\frac{y_t^2}{\sigma^2_t(\boldsymbol{\theta})}).\label{eq:gl2}
\end{equation}
The score-function for the primary parameters of the GARCH model is expression (\ref{eq:garch_s}),where $a$ and $b$ are primary parameters:
\begin{equation}
    \boldsymbol{s} = \left(\frac{\partial {\rm log}(L)}{\partial a},\frac{\partial {\rm log}(L)}{\partial b}\right)^T.    \label{eq:garch_s}
\end{equation}
The value of $\boldsymbol{V}_{\omega}/m$ is obtained by the inverse of the per sample FIM of $\omega$, whose expectation is equal to the expectation of the negative Hessian matrix.


\subsection{Neural networks}
Neural networks (NNs) plays a crucial role in the field of machine learning, especially in artificial intelligence. They streamline the modeling process by autonomously extracting intricate patterns and relationships from raw data, constructing multi-level models that can effectively process complex information. Their adaptability, scalability, and versatility render them indispensable in diverse applications, including, but not limited to, image and speech recognition, natural language processing, control systems and automated decision-making. However, the efficient autonomous learning ability of NN models comes with a drawback --- a large number of parameters. The extensive computations required for parameter estimations not only consume time but also contribute to increased greenhouse gas emissions. Caballar \cite{Caballar2024we} suggested that the hardware in data centers and data transmission networks collectively contribute to approximately 1 percent of energy-related greenhouse gas emissions. In response, greener software engineering has emerged as a burgeoning discipline, advocating for best practices aimed at developing applications that minimize carbon footprints.

One strategy within greener AI involves simplifying model architectures by reducing parameters and optimizing hyperparameters—external model controls—using a tuning strategy that minimizes the number of iterations \cite{Caballar2024we}. Pre-defining certain "uninteresting" bias terms ($\boldsymbol{b}$) is an approach to simplify parameter estimation and boost learning efficiency in constructing NN models. Mohan et al. \cite{mohan2019robust} proposed that bias-term-free deep NN can effectively ensure the accuracy of the model, especially for image recognition. The bias-free NN has a good noise reduction effect which even improves the accuracy of model prediction. 

Nonetheless, a critical consideration lies in addressing the uncertainty associated with the bias term. In NNs, let $\boldsymbol{x}$ denote row information (samples) in input layer, $\boldsymbol{y}$ denote the set of required output in final layer, $\boldsymbol{w}$ and $\boldsymbol{b}$ are the true weight vector and true bias term respectively, $\hat{\boldsymbol{w}}$ is the estimated weight vector, viewed as primary parameters, $\hat{\boldsymbol{b}}$ are the estimated bias terms, viewed as nuisance parameters. We set \(\boldsymbol{\mu_{y}}(\boldsymbol{x})\) a the true mean of output given input $\boldsymbol{x}$, and \(\hat{\boldsymbol{\mu}}_{\boldsymbol{y}}(\boldsymbol{x};\hat{\boldsymbol{w}},\hat{\boldsymbol{b}})\) is the estimated mean of output given input $\boldsymbol{x}$. Penny and Roberts \cite{penny1997neural} noted that, for NN models, uncertainty of prediction arises from two sources related to parameters. One source is the presence of multiple local minima in the error function, leading to various potential parameter values. The other source is errors introduced by sub-optimal training, such as those occurring in early termination training algorithms.

Building on these insights, Dybowski and Roberts \cite{dybowski2001confidence} argued that the uncertainty of $\hat{\boldsymbol{\mu}}_{\boldsymbol{y}}(\boldsymbol{x};\hat{\boldsymbol{w}},\hat{\boldsymbol{b}})$ is largely attributed to the uncertainty of parameters. Many researchers also paid more attention to the uncertainty of parameters to improve the accuracy of prediction. Blundell et al. \cite{Blundell2015WeightUncertainty} proposed a novel algorithm, Bayes by Backprop, for NN learning with uncertainty on weights, enhancing predictive capabilities in nonlinear regression. Antoran et al. \cite{Antoran2020Depth} proposed performing probabilistic reasoning over NN depth, which enables uncertainty estimation in deep learning with a single forward pass, and provides calibrated uncertainty estimates. The uncertainties in both primary and nuisance parameters are related to the accuracy of prediction. Consequently, investigating the impact of uncertainty in nuisance parameters holds practical significance for NN models, providing valuable insights for optimizing performance and addressing the complexities associated with parameterization in research studies.


\subsubsection{Model structure}
Let $\boldsymbol{x}$ denote row information (samples) in input layer, $\boldsymbol{y}$ denote the set of NN output in final layer. The general structure of a NN with $K$ hidden layers can be represented as follows structure:

\begin{equation}
\boldsymbol{z}^{(1)} = \sigma(\boldsymbol{x} \cdot \boldsymbol{w}^{(1)} + \boldsymbol{b}^{(1)}),
\end{equation}
\begin{equation}
\boldsymbol{z}^{(k)} = \sigma(\boldsymbol{z}^{(k-1)} \cdot \boldsymbol{w}^{(k)} + \boldsymbol{b}^{(k)}),
\end{equation}
\begin{equation}
\boldsymbol{y} = \sigma(\boldsymbol{z}^{(K)} \cdot \boldsymbol{w}^{(K+1)} + \boldsymbol{b}^{(K+1)}), 
\end{equation}
where $ \boldsymbol{z}^{(k)}$ is the elements on hidden layer $k$ and $\sigma(\cdot) $ is the activation function e.g., sigmoid or ReLU. 
For example, the structure of NN shows as follows when $K$=2:
\begin{figure}[H]
\centering
\includegraphics[ width=0.6\textwidth]{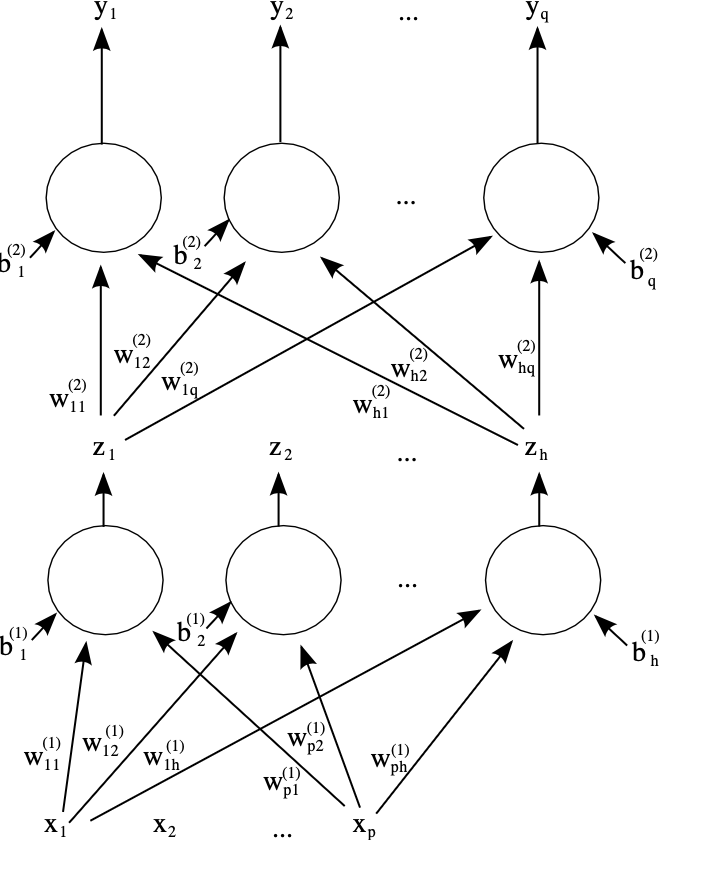} 
\captionsetup{font=small}
\caption{The structure of a NN with 2 hidden layers.}
\label{Fig.nn} 
\end{figure}
Following the bias-free-term deep NN \cite{mohan2019robust}, $\boldsymbol{b}$ are viewed as nuisance parameters who is given as 0 in our paper. A common loss function used in training is the mean squared error (MSE), defined as expression (\ref{eq:loss_function}):
\begin{equation}
C(\boldsymbol{y}, \hat{\boldsymbol{y}}) = \frac{1}{n} \sum_{i=1}^{n} L(y_i,\hat{\boldsymbol{y}}_i)\label{eq:loss_function}
\end{equation}
where $\boldsymbol{y}$ is the true output (true label), $\hat{\boldsymbol{y}}$ is the predicted output, $n$ is the number of samples and $L(\cdot)$ is the function used to calculate loss. In practice, $L_1=|\boldsymbol{y}-\hat{\boldsymbol{y}}|$ and $L_2=(\boldsymbol{y}-\hat{\boldsymbol{y}})^2$ are commonly used as $L(\cdot)$. And the method used to estimate parameters is backpropagation, which calculates the derivative of the loss function and utilizes the chain rule and gradient descent to find optimized point.

If the loss function in mean-squared error in $L_2$ form is used, the estimated parameter vector is the maximum-likelihood estimate (Dybowski and Roberts \cite{dybowski2001confidence}) when the distribution of $\boldsymbol{y}$ about $\boldsymbol{\mu}_{\boldsymbol{y}}(\boldsymbol{x})$ is assumed to be Gaussian and $\boldsymbol{y}$ is $d\times1$ vector:
\begin{equation}
    f(\boldsymbol{y}|\boldsymbol{x},\boldsymbol{w},\boldsymbol{b}) = (2\pi)^{-\frac{d}{2}} |{\rm det}(\boldsymbol{\Sigma_{y}})|^{-\frac{1}{2}}\cdot {\rm exp}\left(-\frac{1}{2}(\boldsymbol{\mu}_{\boldsymbol{y}}(\boldsymbol{x}))-\boldsymbol{y})^{\rm T} \boldsymbol{\Sigma_{y}}^{-1} (\boldsymbol{\mu}_{\boldsymbol{y}}(\boldsymbol{x}))-\boldsymbol{y})\right).\label{eq:nn-norm}
\end{equation}

The MLE is as expression (\ref{eq:nn-mle}):
\begin{equation}
    \hat{\boldsymbol{w}}_{MLE}={\rm arg_w min} Err(\boldsymbol{w}),\label{eq:nn-mle}
\end{equation}
\begin{equation}
Err(\boldsymbol{w})= \frac{1}{n} \sum_{i=1}^{n} (\hat{\boldsymbol{\mu}}_{\boldsymbol{y}}(\boldsymbol{x}^{(i)};\hat{\boldsymbol{w}},\hat{\boldsymbol{b}}) - \boldsymbol{y})^{\rm T}(\hat{\boldsymbol{\mu}}_{\boldsymbol{y}}(\boldsymbol{x}^{(i)};\hat{\boldsymbol{w}},\hat{\boldsymbol{b}}) - \boldsymbol{y}).\label{eq:nn-mle-1}
\end{equation}


Recalling the general nuisance parameter theory in expression (\ref{theorem}), the adjusted variance of NN is as expression (\ref{eq:NN_var}):
\begin{equation}
    {\rm{cov}}(\hat{\boldsymbol{w}}|\boldsymbol{x}^{(n)},\hat{\boldsymbol{b}}^{(m)})=\frac{\boldsymbol{V}_{\boldsymbol{w}}}{n}+\boldsymbol{D}_1\frac{\boldsymbol{V}_{\boldsymbol{b}}}{m}\boldsymbol{D}_1^T,\label{eq:NN_var}
\end{equation}
where $\boldsymbol{V}_{\boldsymbol{w}}/n = {\rm cov}(\hat{\boldsymbol{w}}|\boldsymbol{x},\boldsymbol{b}^*)$, $\boldsymbol{V}_{\boldsymbol{b}}/m$ is the variance matrix of bias term, $\nabla C$ is the score function ($\boldsymbol{s}$), and $\boldsymbol{D}_1=-\left(\partial \nabla C/\partial \boldsymbol{w}^T\right)^{-1}\left(\partial \nabla C / \partial \boldsymbol{b}^T\right)$ .

Due to the extensive number of parameters in the NN model, it becomes impractical to visualize the impact of the uncertainty of nuisance parameters on each primary parameter. To address this challenge, this paper incorporates the concept of 90\% confidence intervals of each of the components of $\hat{\boldsymbol{\mu}}_{\boldsymbol{y}}(\boldsymbol{x};\hat{\boldsymbol{w}},\hat{\boldsymbol{b}})$ assuming a Gaussian target noise distribution (Dybowski and Roberts \cite{dybowski2001confidence}) into the analysis. The 90\% confidence intervals of $\hat{\mu}_{y_i}(\boldsymbol{x};\hat{\boldsymbol{w}},\hat{\boldsymbol{b}})$, the component of vector $\hat{\boldsymbol{\mu}}_{\boldsymbol{y}}(\boldsymbol{x};\hat{\boldsymbol{w}},\hat{\boldsymbol{b}})$ associated with $y_i$, is expressed as expression (\ref{eq:ci}),  By comparing the confidence intervals of predictions with adjusted variance (due to nuisance parameters) and unadjusted variance, the effect of nuisance parameter uncertainty is illustrated. This method helps make results visualized and integrates the impact of uncertainty in nuisance parameters on the overall model.

\begin{equation}
\hat{\mu}_{y_i}(\boldsymbol{x};\hat{\boldsymbol{w}},\hat{\boldsymbol{b}}) \pm \boldsymbol{z}_{0.025} \sqrt{ \boldsymbol{g}^T(\boldsymbol{x}) \boldsymbol{\Sigma} \boldsymbol{g}(\boldsymbol{x})},\label{eq:ci}
\end{equation} 
where vector $\boldsymbol{g}(\boldsymbol{x})$ is the partial derivative $\partial\hat{\mu}_{y_i}(\boldsymbol{x};\hat{\boldsymbol{w}},\hat{\boldsymbol{b}})/\partial \hat{\boldsymbol{w}}$ and $\boldsymbol{\Sigma}$ is covariance matrix for $\hat{\boldsymbol{w}}$.


In NN, it is easy to get vector $\partial C/\partial \hat{\boldsymbol{w} }$ during backpropragation. Regarding the scalar $\partial\hat{\mu}_{y_i}(\boldsymbol{x};\hat{\boldsymbol{w}},\hat{\boldsymbol{b}})/\partial C$, the loss function is first stored in a variable, which is referenced in subsequent steps in model. Then small adjustments are made to this variable for the numerical calculations. Thus, we get vector $\partial\hat{\mu}_{y_i}
(\boldsymbol{x};\hat{\boldsymbol{w}},\hat{\boldsymbol{b}})/\partial \hat{\boldsymbol{w}}$ as expression (\ref{eq:nn_mu3}) below:
\begin{equation}
\frac{\partial\hat{\mu}_{y_i}(\boldsymbol{x};\hat{\boldsymbol{w}},\hat{\boldsymbol{b}})}{\partial\hat{\boldsymbol{w} }}=\frac{\partial\hat{\mu}_{y_i}(\boldsymbol{x};\hat{\boldsymbol{w}},\hat{\boldsymbol{b}})}{\partial C} \cdot\frac{\partial C}{\partial \hat{\boldsymbol{w} }} .\label{eq:nn_mu3}
\end{equation}
Then the 90\% confidence intervals of $\hat{\mu}_{y_i}(\boldsymbol{x};\hat{\boldsymbol{w}},\hat{\boldsymbol{b}})$, the component of vector $\hat{\boldsymbol{\mu}}_{\boldsymbol{y}}(\boldsymbol{x};\hat{\boldsymbol{w}},\hat{\boldsymbol{b}})$ associated with $y_i$, is as expression (\ref{eq:nn_ci2}) below:
\begin{equation}
\hat{\mu}_{y_i}(\boldsymbol{x};\hat{\boldsymbol{w}},\hat{\boldsymbol{b}}) \pm \boldsymbol{z}_{0.025} \sqrt{\left(\frac{\partial\hat{\mu}_{y_i}(\boldsymbol{x};\hat{\boldsymbol{w}},\hat{\boldsymbol{b}})}{\partial C} \cdot \frac{\partial C}{\partial \hat{\boldsymbol{w}}}\right)^T \boldsymbol{\Sigma}  \frac{\partial\hat{\mu}_{y_i}(\boldsymbol{x};\hat{\boldsymbol{w}},\hat{\boldsymbol{b}})}{\partial C} \cdot\frac{\partial C}{\partial \hat{\boldsymbol{w}}}}.\label{eq:nn_ci2}
\end{equation} 

In comparison between predicted results and actual results in NN, the actual results, when transformed into dummy variables, form a binary array where the actual label corresponding to a set of inputs is 1, and the rest of the labels are 0. This can be considered equivalent to probabilities. Since the NN's selection process entails choosing the label with the highest predicted value as the final result when determining the outcome for a single input sample, we calculate the probability of each label's output value in the overall context. The predicted results for each input sample are turned into an array representing the probabilities of this sample belonging to each label. The two probability arrays given one sample ($\boldsymbol{x}^{(i)}$) are used as $\boldsymbol{y}^{(i)}$ and $\hat{\boldsymbol{\mu}}_{\boldsymbol{y}}(\boldsymbol{x}^{(i)}; \hat{\boldsymbol{w}},\hat{\boldsymbol{b}})$ in confidence interval calculations. And $\hat{\boldsymbol{\mu}}_{\boldsymbol{y}}(\boldsymbol{x};\hat{\boldsymbol{w}},\hat{\boldsymbol{b}})$ represents the sample mean of probabilities for each class (label) across the entire input data set.


\section{Numerical Studies}

\subsection{GARCH Model Simulation}

Consider the GARCH setting of Section 4.1. To better capture the effect of the adjusted variance in representing the uncertainty of the nuisance parameter ($\alpha = w $), two sets of mutually independent sample datasets with a sample size of 1000 each are simulated under a fixed-parameter GARCH model.
One dataset (Sample 1) is selected to estimate the primary parameters ($\hat{\boldsymbol{\theta}} = (a,b)^T$) given the nuisance parameter value ($\hat{\alpha}$) and obtain ${\rm cov}(\hat{\boldsymbol{\theta}}|\hat{\alpha})$ (representing $\boldsymbol{V_\theta}/n$) in the conventional way—by inverting its unit sample FIM. 
$\boldsymbol{V}_{\alpha}/m$ is obtained by estimating FIM of nuisance parameter from the other dataset (Sample 2).
As in the simulation process, the true value of nuisance parameters ($\alpha^*$) is known, then $\boldsymbol{V}_{\boldsymbol{\theta}}$ is estimated from Sample 1 given $\alpha^*$. 
In empirical research, since $\alpha^*$ is unknown, and $\hat{\alpha}$ given during the estimation of the primary parameters is usually the value closest to $\alpha^*$ based on subjective knowledge, the $\rm{cov}(\hat{\boldsymbol{\theta}}|\hat{\alpha})$ obtained from Sample 1 could be directly used instead.

We selected two cases to reflect the simulation effect, plotted their 90\% confidence interval ellipses of primary parameters, and identified the estimated primary parameter $\hat{\boldsymbol{\theta}}$ and true value of the primary parameter $\boldsymbol{\theta}^*$. 

\begin{figure}[H]
\centering
\includegraphics[ width=0.8\textwidth]{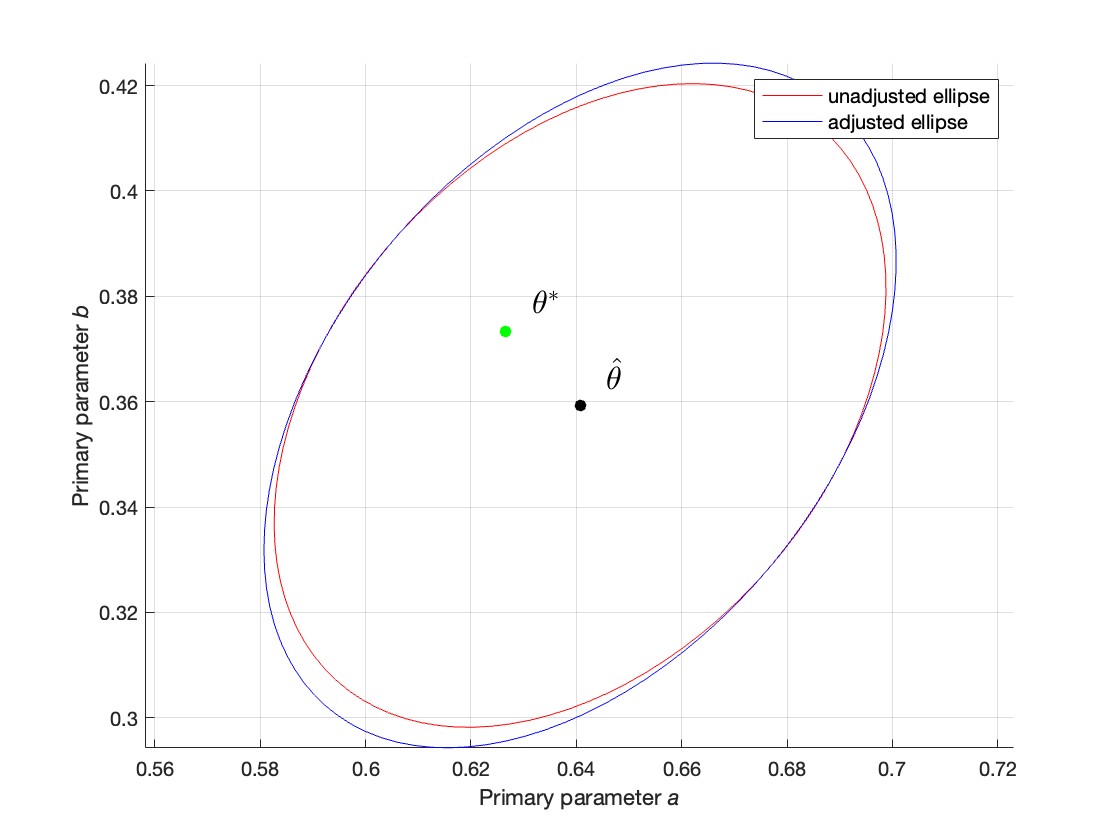} 
\captionsetup{font=small}
\caption{90\% confidence ellipses for primary parameters $\hat {\boldsymbol{\theta}}$ when the relative error of $\hat {\alpha}^{(m)}$ to $\alpha^*$ is 5\%.}
\label{Fig.case1} 
\end{figure}

In case 1, we set the relative error between $\hat{\alpha}$ and $\alpha^*$ to be 5\% ($\hat{\alpha}=0.95\alpha^*$). The result is shown in Figure \ref{Fig.case1}, where the ellipses are based on $\boldsymbol{V}_{\boldsymbol{\theta}}/n$ (unadjusted covariance matrix) and (\ref{eq:garch_var}) (adjusted covariance matrix).

As shown in Figure \ref{Fig.case1}, when $\hat{\alpha}$ is very close to $\alpha^*$, the adjusted variance is close to the unadjusted variance. And both 90\% ellipses based on the adjusted variance and unadjusted variance cover $\boldsymbol{\theta}^*$.

In case 2, on the sample of the same dataset and model as case 1, $\hat{\alpha}$ is changed so that the relative error between it and $\alpha^*$ is 30\% of the value of $\alpha^*$ and the estimation error of $\hat{\boldsymbol{\theta}}$ becomes larger ($\hat{\alpha}=0.7\alpha^*$). The result is shown in Figure \ref{Fig.case2}:

\begin{figure}[H]
\centering
\includegraphics[ width=0.8\textwidth]{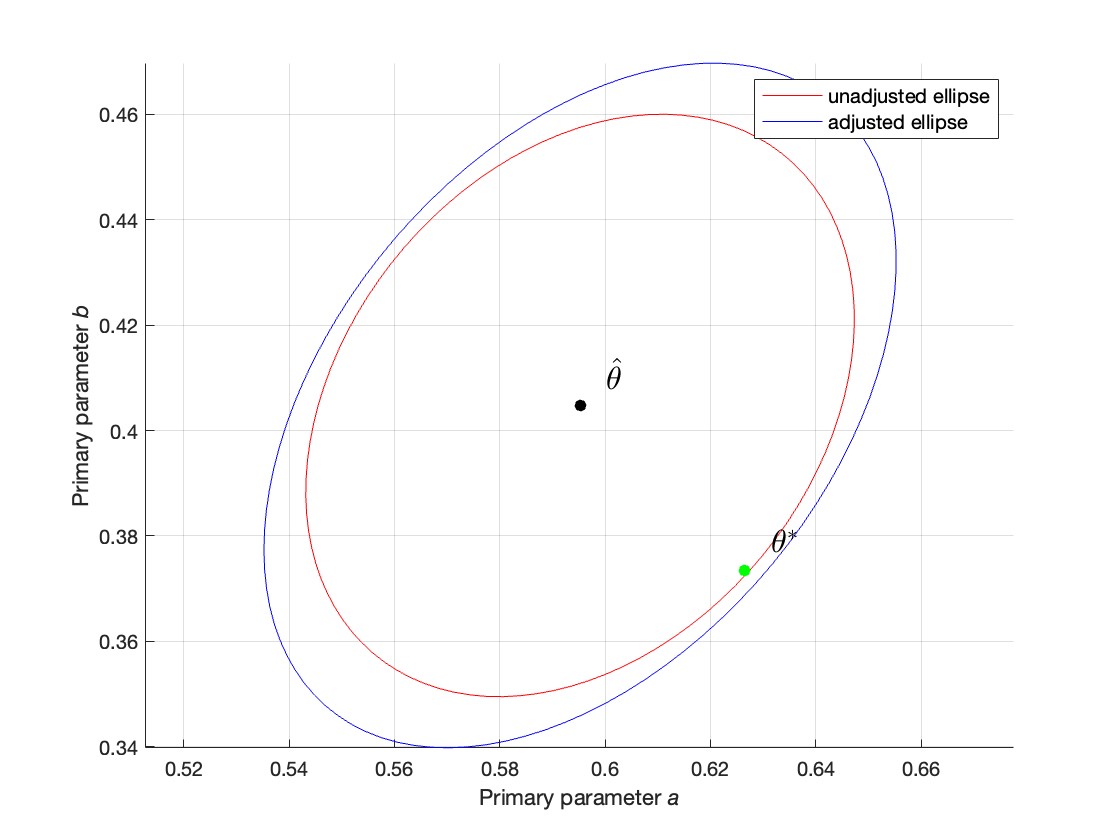} 
\captionsetup{font=small}
\caption{90\% confidence ellipses for primary parameters $\hat {\boldsymbol{\theta}}$ when the relative error of $\hat {\alpha}^{(m)}$ to $\alpha^*$ is 30\%.}
\label{Fig.case2} 
\end{figure}

In this case, the 90\% ellipse based on the adjusted variance covers $\boldsymbol{\theta}^*$ while the unadjusted ellipse almost does not. Also, comparing it with case 1, it is observed that the adjusted variance is sensitive to the $\hat{\alpha}$ error and becomes larger as the error increases.


In conclusion, drawing from the aforementioned findings, it can be inferred that the 90\% ellipse based on the adjusted variance becomes larger than the ellipse from the unadjusted variance with larger biased error variance of $\hat{\alpha}$ the adjusted ellipse is more likely to cover the true value of primary parameters at the specified probability level.


\subsection{GARCH model empirical research}
EUR/USD (the exchange rate between the Euro and the United States Dollar) and GBP/USD (the exchange rate between the British Pound and the United States Dollar) are standard and widely accepted pairs in the financial market. The Eurozone and the United Kingdom, as major economic regions, share common influences from global economic factors, leading to similarities in volatility patterns in currency pairs. Both currencies are sensitive to global risk factors, with movements responding similarly to changes in economic uncertainties, geopolitical events, and shifts in market sentiment. Moreover, the use of the US Dollar (USD) as the quote currency in both pairs contributes to analogous GARCH patterns, as changes in the USD impact the volatility of EUR/USD and GBP/USD alike. Moreover, on one hand, trade relationships and economic ties exist between the Eurozone and the United Kingdom. On the other hand, traders and investors often participate in both EUR/USD and GBP/USD markets, causing them to have correlated market participants with similar trading strategies. These factors contribute to the synchronization of volatility patterns between these currency pairs. Therefore, we assume that, without externality, the pairs should have the same GARCH model.

Additionally, log-linear realized GARCH, which utilizes realized volatility as a sample and logarithmically transforms realized volatility to estimate parameters in the GARCH model, provides a better empirical GARCH prediction without externality due to the additional information in realized measurement and the removal of heteroscedasticity (Hansen et al. \cite{hansen2012realized}).

The dataset is the realized volatility of EUR/USD ($RV_{EUR/USD}$) and GBP/USD ($RV_{GBP/USD}$) from January 2, 2020, to November 8, 2023. 

\begin{figure}[H]
\centering
\includegraphics[ width=0.8\textwidth]{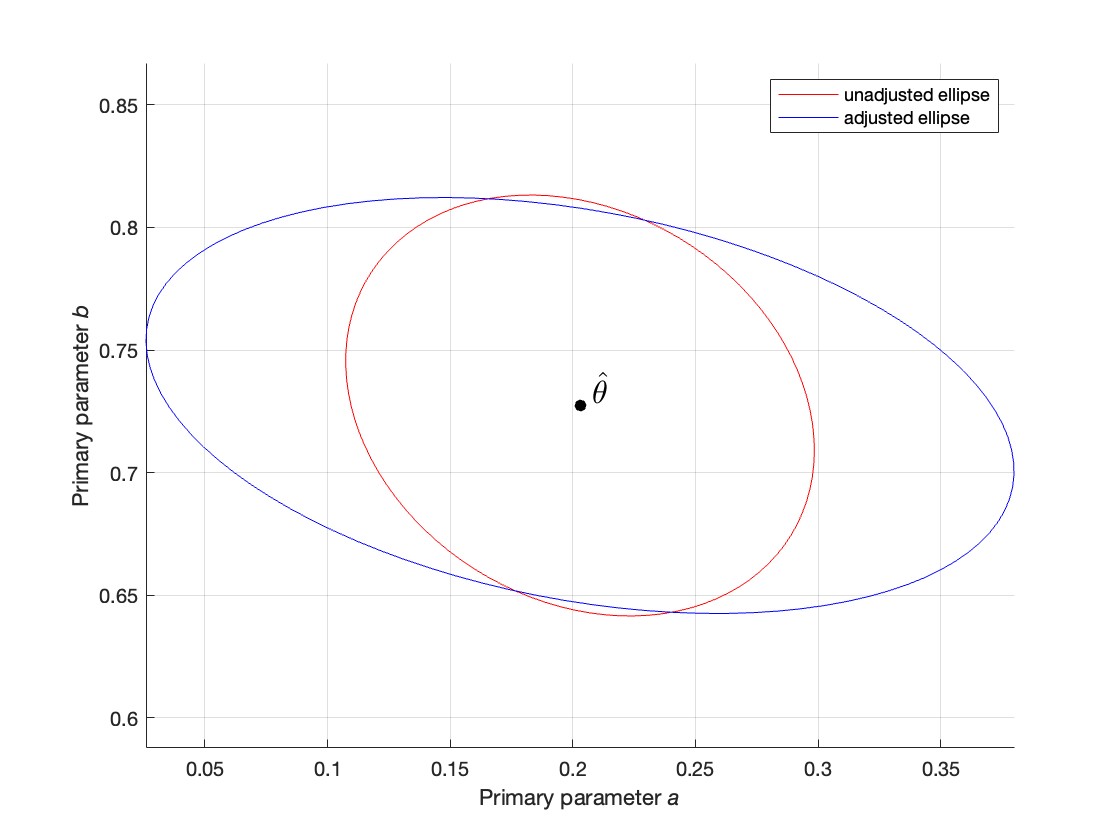} 
\captionsetup{font=small}
\caption{90\% confidence ellipses for primary parameters $\hat {\boldsymbol{\theta}}$ in EUR/USD and GBP/USD pair.}
\label{Fig.nui_nan} 
\end{figure}

We establish a GARCH model for log$(RV_{EUR/USD})$, taking 
log($RV_{GBP/USD}$) as another independent data to estimate $\boldsymbol{V}_{\alpha}/m$.
Usually, for $\hat{\alpha}$, the researcher will assign the value closest to $\alpha^*$ according to subjective knowledge. And, as another independent data from significant same models, the parameters of log($RV_{GBP/USD}$) should be significantly same in theory. 
Therefore, $\alpha$ for log($RV_{GBP/USD}$) obtained during the estimation of $\boldsymbol{V}_{\alpha}/m$ is used as $\hat{\alpha}$ to estimate the primary parameter of the GARCH model for log$(RV_{EUR/USD})$ . 
Also, we draw the 90\% confidence ellipse for the primary parameter, marking the estimated primary parameter on the plot. The results are depicted in Figure \ref{Fig.nui_nan}.

The adjusted ellipse proves to be larger than the unadjusted ellipse, nearly enclosing the latter, which shows the adjusted variance effectively includes the uncertainty of the nuisance parameter. 

\subsection{Neural network empirical research }
We follow the notation and concepts discussed in Section 4.2. In the experimental setup, the fundamental variance of primary parameters (\(\boldsymbol{V_w}/n\)) in the NN model, is estimated by computing the inverse of the FIM per sample. This is facilitated by the fact that the NN model is optimized using gradient descent, which provides the gradient of the loss function with respect to the parameters.  $-(\partial \nabla C / \partial \boldsymbol{w}^T)^{-1}$, the inverse of negative Hessian matrix, is estimated as the inverse of the FIM for primary parameters of model trained on the test set. For the computation of \(\partial \nabla C / \partial \boldsymbol{b}^T\), a two-sided numerical finite-difference method is employed. This method involves calculating the average impact of the smallest increase and decrease in each bias term on the gradient of the loss function for each weight term, and combining them as required matrix.

Estimating the nuisance parameters covariance matrix \(\boldsymbol{V_b}/m\), is more intricate due to the NN having a considerably larger number of primary parameters compared to nuisance parameters. It is inaccurate to directly estimate the variance through the nuisance parameter component from the FIM given primary parameters. Pearce et al. \cite{pearce2020uncertainty} confirmed, through randomized Maximum a Posteriori (MAP) Sampling and Root Mean Square (RMS), that the conditional parameter likelihoods of common datasets, such as MNIST, for actual NN adheres to an approximate normal distribution. Building upon this insight, we randomly selected multiple data groups from the test set, independent of the training set, to estimate the model. A set of estimated bias parameters was obtained from each data group, forming a sample group for bias parameter estimation.

We adopt an approach for estimating \(\boldsymbol{V_b}/m\): First, assuming a prior normal distribution, we estimate the variance of each bias parameter using Bayesian inference. Second, we compute the statistical covariance matrix for all sample bias data groups, and observing that the covariance between different bias terms is nearly 0. Then, for this NN trained with gradient descent to minimize the \(L_2\) loss, we posit that the bias parameters follow a Gaussian distribution \(\boldsymbol{b} \sim N(\boldsymbol{0}, \tau^2\boldsymbol{I})\), as in Basri et al. \cite{basri2020frequency}. This yields a diagonal matrix, with each diagonal value representing the estimated variance of one bias parameter and the remaining values set to 0, serving as our \(\boldsymbol{V_b}/m\).

In the selection of research data, adhering to the assumption that the distribution of \(\boldsymbol{y}\) about \(\boldsymbol{\mu_{y}}(\boldsymbol{x})\) is Gaussian, MNIST (Modified National Institute of Standards and Technology), a well-known grayscale images dataset of handwritten digits from 0 through 9, is chosen as the research dataset. Developed by LeCun et al. \cite{lecun1998gradient}, the MNIST database includes both training and test sets, ensuring independence as no author is involved in both sets. Additionally, each sample within a number group is written by a different author, ensuring independence within the group. With a large dataset size of 70,000 instances, the MNIST dataset aligns well with our assumptions.

Given MNIST's 28$\times$28 pixel black and white images, where each sample's input size is 784, if the input layer has dimensions 784$\times$16, this results in 12,544 parameters. Computing the inverse of the FIM for such a large number of parameters is computationally intensive. Recognizing that the primary function of the input layer is to extract initial image information and subsequent layers refine predictions, our focus is on the practical study of primary parameters excluding the input layer.

\begin{figure}[H]
\centering
\includegraphics[ width=0.7\textwidth]{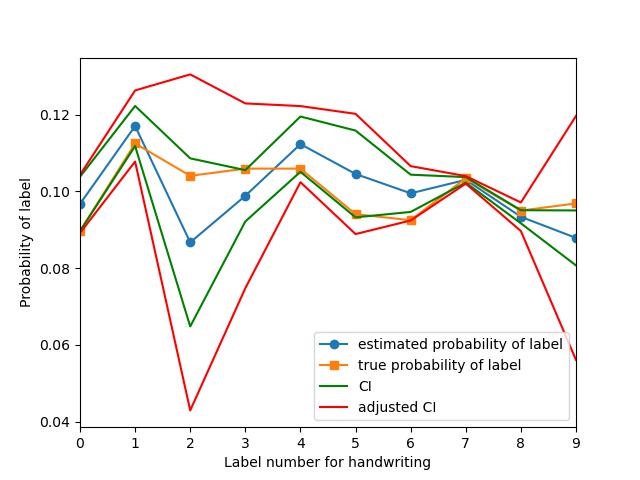} 
\captionsetup{font=small}
\caption{Case 1: 90\% Confidence interval (CI) of each label's probability for NN with 2 hidden layers.}
\label{Fig.NNcase1} 
\end{figure}

\begin{figure}[H]
\centering
\includegraphics[ width=0.7\textwidth]{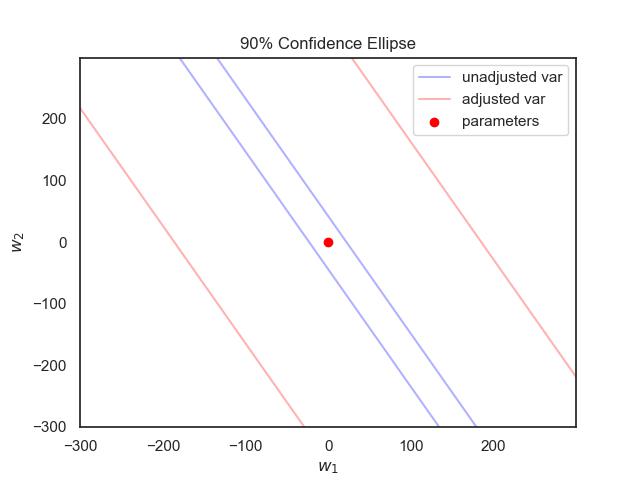} 
\captionsetup{font=small}
\caption{Case 1: Small part of 90\% confidence ellipses of two parameters ($w_1,w_2$) for NN with 2 hidden layers.}
\label{Fig.NNcase1-1} 
\end{figure}

\begin{figure}[H]
\centering
\includegraphics[ width=0.7\textwidth]{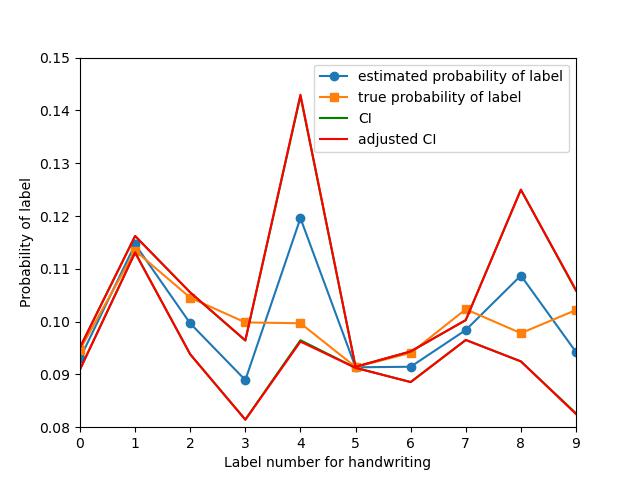} 
\captionsetup{font=small}
\caption{Case 2: 90\% Confidence interval (CI) of each label's probability for NN with 4 hidden layers. Visually, the adjusted and unadjusted CIs are identical.}
\label{Fig.NNcase2} 
\end{figure}

\begin{figure}[H]
\centering
\includegraphics[ width=0.7\textwidth]{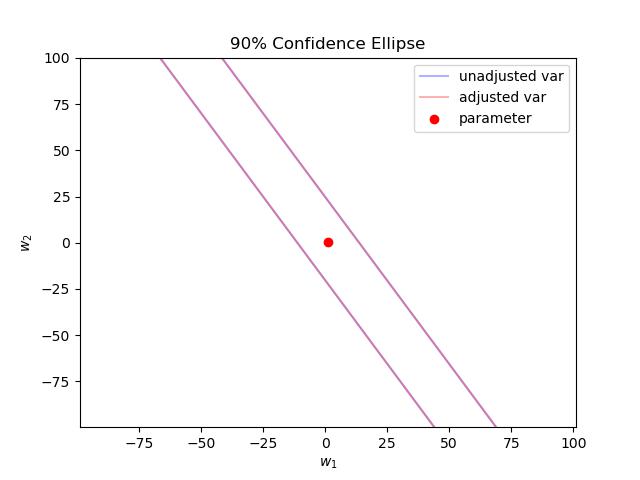} 
\captionsetup{font=small}
\caption{Case 2: Small part of 90\% confidence ellipses of two parameters ($w_1,w_2$) for NN with 4 hidden layers. Visually, the adjusted and unadjusted ellipses are identical.}
\label{Fig.NNcase2-1} 
\end{figure}

Using MNIST, the handwritten digit dataset with 10 labels, we ultimately obtain 10 actual probabilities, 10 predicted probabilities, and their corresponding adjusted and unadjusted 90\% confidence intervals for each label. Subsequently, we visualize the predicted probability along with the adjusted and unadjusted 90\% two-sided confidence intervals, providing a means to observe the impact of uncertainty in nuisance parameters. Simultaneously, any two primary parameters from the final layer of a specific label are selected, and the adjusted and unadjusted variances of these parameters are calculated. A 90\% confidence ellipse is plotted for each. The results from the above are shown in Figures \ref{Fig.NNcase1}-\ref{Fig.NNcase2-1}

More specifically, to further evaluate the impact of uncertainty in nuisance parameters, two NN scenarios are designed, and all models are trained to achieve above 90\% accuracy to mitigate the impact of accuracy variations. In case 1, the NN model follows a simple two-layer hidden layer structure with 16 neurons in each hidden layer, resulting in 416 primary parameters of interest and 42 nuisance parameters. The results are shown in Figure \ref{Fig.NNcase1}. Also, the 90\% confidence ellipse of the first and second parameters of the output transform function from label number 1 are selected as a sample case to show in Figure \ref{Fig.NNcase1-1}.

From Figures \ref{Fig.NNcase1} and \ref{Fig.NNcase1-1}, it is noted a significant expansion in the adjusted confidence interval compared to the unadjusted confidence interval, effectively encompassing the true label probabilities. Furthermore, when examining individual pairs of primary parameters, as in Figure \ref{Fig.NNcase1-1}, the adjusted variance ellipse proves to be larger than the unadjusted variance ellipse. This leads to the conclusion that, for this simple NN structure, the uncertainty in nuisance parameters holds substantial influence, and the adjusted variance is more likely to encompass the true results.

Moving on to case 2, we construct a deeper NN with four hidden layers, each still comprising 16 neurons. This configuration results in 928 primary parameters of interest and 74 nuisance parameters. The 90\% confidence ellipse of the first and second parameters of the output transform function from label number 1 are selected as a sample case to show. The results are shown in Figures \ref{Fig.NNcase2} and \ref{Fig.NNcase2-1}.

From Figures \ref{Fig.NNcase2} and \ref{Fig.NNcase2-1}, it is found that the adjusted and unadjusted confidence intervals nearly overlap. Likewise, when examining individual pairs of primary parameters (Figure \ref{Fig.NNcase2-1}), the adjusted variance ellipse and unadjusted variance ellipse also exhibit substantial overlap. This leads to the conclusion that, for a deeper NN with more layers, the uncertainty in nuisance parameters has minimal impact on predictive performance. Notably, this experimental result aligns with the conclusion drawn by Mohan et al. \cite{mohan2019robust} regarding bias-free deep NN's characteristics.

\section{Conclusion}
For method's performance,  we confirm that the accuracy of our adjusted variances' statistical inferences is comparable to those obtained from estimating all parameters by simulating a numerical test of an exponential model.

In application, for the GARCH model, we illustrate that the confidence ellipse of the estimated primary parameters, based on adjusted variance due to nuisance parameters, is more likely to encompass the true values of primary parameters than the unadjusted confidence ellipse.

For the NN, numerical investigations indicate that the effect of uncertainty in nuisance parameters, specifically the bias terms, is correlated with the depth of the model's structure. As the depth, represented by the number of hidden layers, increases, the impact of bias uncertainty diminishes. This underscores the resilience of deep NN with a bias-free structure in maintaining high-quality predictions (Mohan et al. \cite{mohan2019robust}). This also substantiates a common observation among diverse scholars. Namely, the prerequisite they posit for training models to mitigate uncertainty stemming from various error sources is consistently rooted in the utilization of deep NN (Kim et al. \cite{kim2019learning}, Adeli and Zhao \cite{adeli2020biasresilient} and Rahaman et al. \cite{rahaman2019}). Additionally, in situations where uncertainty has a discernible effect, the 90\% confidence interval for predicted results, based on the nuisance parameters adjustment, is more likely to contain the true outcomes.

The above asymptotically based adjustment of confidence regions due to nuisance parameters has been successfully applied in the space-state models, the GARCH models, and NNs giving credence to the value of the method across different data science models. This method not only allows for reduced parameterization in primary parameters, its result also helps in the calculation of credible confidence regions for practical applications in the common situation of having uncertain nuisance parameters from other sources. 

\end{spacing}
\newpage
\bibliographystyle{ieeetr}
\bibliography{main.bib}
\end{document}